# Unintentional F doping of the surface of SrTiO$_3$(001) etched in HF acid – structure and electronic properties

S.A. Chambers[a,*], T.C. Droubay[a], C. Capan[b], G.Y. Sun[c]

[a]*Computational and Fundamental Sciences Directorate, Pacific Northwest National Laboratory, Richland, WA*
[b]*Department of Physics, Washington State University, Pullman, WA*
[c]*Department of Materials Science and Engineering, Nanjing University of Science and Technology, People's Republic of China*

**Abstract**

We show that the HF acid etch commonly used to prepare SrTiO$_3$(001) for heteroepitaxial growth of complex oxides results in a non-negligible level of F doping within the terminal surface layer of TiO$_2$. Using a combination of x-ray photoelectron spectroscopy and scanned angle x-ray photoelectron diffraction, we determine that on average ~13 % of the O anions in the surface layer are replaced by F, but that F does not occupy O sites in deeper layers. Despite this perturbation to the surface, the Fermi level remains unpinned, and the surface-state density, which determines the amount of band bending, is driven by factors other than F doping. The presence of F at the STO surface is expected to result in lower electron mobilities at complex oxide heterojunctions involving STO substrates because of impurity scattering. Unintentional F doping can be substantially reduced by replacing the HF-etch step with a boil in deionized water, which in conjunction with an oxygen tube furnace anneal, leaves the surface flat and TiO$_2$ terminated.

PACS: 68.47.Gh, 79.60.-i, 61.05.js

*Tel: +1 509 371 6517; fax +1 509 371 6066.
*E-mail address*: sa.chambers@pnnl.gov

## 1. Introduction

SrTiO$_3$(001) (STO) is a commonly used substrate for oxide heteroepitaxy. Moreover, when doped with either Nb, La, or oxygen vacancies, STO is a potentially useful *n*-type semiconductor in complex oxide heterostructures. When cleaved in ultrahigh vacuum (UHV), STO(001) has been shown to exhibit interesting electronic structure properties that have been interpreted as resulting from two-dimensional electron gas formation on the surface [1, 2]. However, UHV cleaving is not a practical way to generate substrates for heteroepitaxy and heterojunction formation. It is important to have a reliable way other than cleaving to prepare STO(001) surfaces that are both atomically flat and free of impurities and defects.

As received from the supplier, STO(001) is highly disordered from polishing and typically does not have a unique surface termination. As a result, it is common practice to soak the STO in water and etch in buffered HF [3], and then anneal at high temperature in flowing O$_2$ [4]. The first two steps preferentially dissolve the SrO terraces, leaving the surface as a roughened array of discontinuous TiO$_2$ mesoscale islands. Annealing in O$_2$ results in mass transport and coalescence of the TiO$_2$ islands into large, flat terraces separated by steps of height equal to ~4 Å, which is close to the lattice parameter of STO (3.91 Å). Although the etched and annealed surface is flat, cathodoluminescence measurements reveal significant defect densities as a result of the HF etch that can be minimized by substituting HCL/HNO$_3$ for HF [5]. It is commonly assumed that following the HF etch, the surface is free of impurities other than carbon. Indeed, there are an abundance of studies reporting electrical and magneto-transport properties for various complex oxide heterostructures involving STO in which no surface analysis other than atomic force microscopy (AFM) was carried out on the STO substrate prior to heterojunction preparation. Moreover, in the relatively sparse literature in which STO surface preparation is discussed, there is to the best of our knowledge no mention of F impurities as a result of the HF etch. In this paper, we use a combination of core-level (CL) and valence band (VB) x-ray photoelectron spectroscopy (XPS), along with x-ray photoelectron diffraction (XPD), to show that: (i) HF etching results in F doping of the surface layer without pinning the Fermi level, and, (ii) the F is not removed by either oxygen plasma treatment at room temperature or annealing at ~550-600$^{o}$C in either high vacuum or oxygen plasma.

## 2. Experimental details

Nb-doped and undoped STO(001) single crystals were etched for 30 sec in a 5% buffered HF solution and rinsed in deionized water. Samples were then annealed in flowing air at 950$^{o}$C for 8 hours in a tube furnace. We have found that the longer anneal results in flatter terraces than does a more conventional one-hour anneal, especially for surfaces with low areal step density [6]. Upon loading into a high vacuum environment, the surfaces were exposed to either a beam of O atoms from an electron cyclotron resonance (ECR) plasma source, or an $O_3$ beam, until all adventitious carbon was removed, as judged by XPS. Some of the surfaces were subsequently annealed either in high vacuum or in a oxygen ECR plasma at 550 – 575$^{o}$C.

All XPS and XPD measurements were carried out using a Gamma Data/Scienta SES200 electron analyzer and monochromatic AlKα x-ray source. The energy resolution for the high-resolution CL and VB spectra was ~0.5 eV. The binding energy scale was calibrated against the Au $4f_{7/2}$ peak at 84.00 ± 0.02 eV and the Au Fermi level at 0.00 ± 0.02 eV using an Au foil. Binding energies for the *n*-STO crystals (which do not charge in XPS) are meaningful in an absolute sense and are reported relative to the Fermi level, allowing us to determine band bending at the surface. Undoped STO samples required the use of a low-energy electron flood gun to eliminate the deleterious effects of surface charging during XPS. As a result, their measured binding energies are not correct on an absolute scale and all spectra for these specimens were shifted to align with those of the Nb-doped samples. The XPD measurements were made with an angular resolution of ~14° (full angle of acceptance) in both polar and azimuthal directions.

## 3. Results and discussion

Fig. 1 shows survey spectra for STO(001) substrates obtained at a take-off angle (θ) of 90°, at which the probe depth ranges from ~5 nm at the VB edge to ~4 nm at the F 1s binding energy. Here we define the probe depth to be ~3 times the electron attenuation length, and ~95% of the measured signal originates within this depth. The top spectrum is for a sample that exhibited the highest amount of F of all those measured, ~6 % of the anions within the probe depth, based on the atomic photoemission cross sections for F 1s and O 1s [7]. The middle spectrum represents the average amount of F measured over all samples (~3 % of the anions), and the bottom spectrum is for a sample that was boiled in deionized water (DI $H_2O$) [8], rather than being etched in HF. Although no F 1s feature is detectable above background in the bottom survey spectrum, it is clearly seen in the high-energy-resolution spectrum, which is shown as an inset. Based on this spectrum and an O 1s spectrum measured under the same instrumental settings, we estimate the F concentration to be ~0.5 % of the anions within the probe depth. These spectra reveal the high level of sensitivity we have to F using XPS with monochromatic x-rays, for which the background is quite low.

In Fig. 2 we show Ti 2p, Sr 3d, O 1s and F 1s high-energy-resolution CL spectra measured at take-off angles of 90° and 10° for a sample exhibiting the average amount of F after cleaning in a oxygen plasma and vacuum annealing for 10 minutes at 575°C. At θ = 10° the probe depth is reduced by a factor of ~5 relative to analysis at θ = 90°. The Ti 2p line shape (Fig. 2a) is the same at both angles, and is characteristic of fully oxidized Ti(IV). There is no evidence for Ti(III), which would be expected if the preparation resulted in the creation of O vacancies. In contrast, there is a significant change in line shape for the Sr 3d peak (Fig. 2b). At θ = 10°, higher-binding-energy shoulders appear on both principal spin-orbit peaks. This manifold can be fitted using the same peak parameters for the principal spin-orbit peaks as are used to fit those in the 90° spectrum, plus two additional peaks with binding energy shifts of ~0.8 and ~1.0 eV relative to the $j = 5/2$ and $3/2$ peaks, as seen in Fig. 3. These additional features have been observed earlier and were interpreted as arising from the presence of $SrO_x$ crystallites on the STO(001) surface that formed as a result of excess Sr in the bulk [9, 10]. However, with F strongly bound to the surface, at least one of the two metal cations should exhibit a higher binding energy than that measured in pure STO, as discussed in more detail below.

Returning to Fig. 2c, the F 1s peak is largely the same at the two angles, and the binding energy is close to that measured for $SrF_2$ [11]. The O 1s spectrum (Fig. 2d) shows a very slight OH-derived peak at θ = 10°, indicating the effectiveness of the combination of oxygen plasma cleaning and vacuum annealing in ridding the surface of atmospheric contaminants. There is no change in the lattice O peak as a result of going to the more surface-sensitive take-off angle.

To determine where F is located within the near-surface region, we turn to XPD results [12]. The angular distributions for dopant atoms are highly sensitive to the local structural environment for these species and their positions relative to the surface [13-18]. We show in Fig. 4a polar scans of the F 1s and O 1s intensities measured in the (100) azimuthal plane. The O 1s scan shows diffraction modulation characteristic of photoemission from anion sites in the perovskite lattice. Specifically, there is a strong forward scattering peak at $\theta = 90°$ resulting from O 1s photoelectrons originating in the second layer and being scattered by Ti ion cores in the first layer. This scattering event is illustrated in the right inset in Fig. 4a where we show a cross section of the B-site-containing (100) plane (referred to as $(100)_B$). Within this plane, strong O 1s forward scattering is expected at $\theta = 90°$ because of the presence of Ti along this exit trajectory. Weaker forward scattering is expected at $\theta = 45°$ due to scattering by first-layer O anions in the lattice. The scattering along $\theta = 45°$ within $(100)_B$ is expected to be weaker than that along $\theta = 90°$ because the atomic number of O is less than that of Ti, and because the interatomic distance is larger, as seen in the structural diagram. Similarly, strong forward scattering is expected at $\theta = 45°$ due to O 1s photoelectrons being emitted in the third layer below the surface and being scattered by Sr in the second layer within (100)-oriented atomic planes containing A sites (designated $(100)_A$ in the left inset). Weaker O 1s forward scattering within $(100)_A$ is expected at $\theta = 90°$ due to the presence of more distant O scatterers in the top layer. In addition to these zeroth-order forward scattering events along low-index directions, higher-order interference events also occur at larger scattering angles, giving rise to a complex diffraction signature that is characteristic of the lattice and the de Broglie wavelength of the outgoing photoelectron [12, 19].

Significantly, if F substitutes for O in anion sites in the second and third layers, the F 1s polar scan will show the same diffraction modulation as the O 1s polar scan. However, the F 1s polar profile is largely featureless, as seen in Fig. 4a. This scan shows a monotonic decrease in intensity with increasing take-off angle, which indicates that F is at or near the surface. We thus conclude that F is not at subsurface anion lattice sites. We use a simple inelastic scattering model to estimate the F atomic concentration as a function of depth. In this model, the F 1s and O 1s intensity polar profiles are expressed as

$$I(\theta)_{F1s} \propto \sum_{i=1}^{n} \sigma_{F1s}\, x_{F,i}\rho_i \exp\left(-\frac{d_i}{\lambda_{F1s}\sin\theta}\right) \qquad (1)$$

and

$$I(\theta)_{O1s} \propto \sum_{i=1}^{n} \sigma_{O1s}\, (1-x_{F,i})\rho_i \exp\left(-\frac{d_i}{\lambda_{O1s}\sin\theta}\right). \qquad (2)$$

Here, $\sigma_{F1s}$ and $\sigma_{O1s}$ are the photoemission cross sections, $x_{F,i}$ is the F mole fraction within the anion sublattice of the $i$th layer, $\rho_i$ is the anion number density in the $i$th layer, $d_i$ is the depth of the $i$th layer, and $\lambda_{F1s}$ and $\lambda_{O1s}$ are the electron attenuation lengths. We vary $x_{F,i}$ within the top few layers until optimal agreement is reached between the measured and modeled ratio $I(\theta)_{F1s}/I(\theta)_{O1s}$. The results are shown in Fig. 4b. The best agreement occurs when $x_{F,i}=0.14$ for $i=1$ (i.e., 14% of the anions in the top $TiO_2$ layer are F and the rest are O), and when $x_{F,i}=0$ for $i>1$ (i.e., there is no F below the first layer). The presence (absence) of diffraction modulation in the O (F) 1s polar scan results in some inverted structure in the F 1s/O 1s peak area ratio that is not captured by this simple continuum model. However, the overall behavior is well reproduced by the model for this F atom distribution.

Using this surface stoichiometry and assuming random occupation of O sites by the F impurities, we can apply the binomial theorem to predict the number of second-layer Sr cations bound to one or more F anions in the first layer relative to the number of Sr cations bound only to O within the probe depth at $\theta = 10°$. The former should exhibit a higher Sr 3d binding energy than the latter by virtue of the higher electronegativity of F compared to O. We calculate an expected ratio of 0.31, which is somewhat less than 0.37, the ratio of the combined area of the Sr 3d peaks at 134.3 eV and 136.2 eV to those at 133.5 and 135.2 eV in Fig. 3. This result is consistent with the higher binding-energy peaks being assigned at least in part to second-layer Sr cations bound to a mix of F and O in the first layer, as depicted in the inset in Fig. 3. Interestingly, a second HF etching reduces, but does not eliminate the higher binding-energy-features, while also not resulting in any measurable change in the F concentration. This result suggests that these high-binding energy features may also be due in part to the presence of SrOx crystallites on the surface, as concluded previously based on the solubility of $SrO_x$ in HF [9, 10].

To further corroborate the surface structure derived from the polar scan, we turn to F 1s and O 1s azimuthal scans at $\theta = 10^o$. These data allow us to discriminate between F substitution for O in the top $TiO_2$ layer and (disordered) F-containing secondary phase formation. The latter could result from, for example, the formation of $SrF_2$ nanocrystallites residing atop the terminal $TiO_2$ layer. Specifically, we use the diffraction modulation exhibited by O 1s photoelectrons at $\theta = 10^o$ as a structural fingerprint for anion sites within the terminal $TiO_2$ layer, and compare these modulations with those measured for F 1s over the same angular range. The results are shown in Fig. 5. At this low take-off angle, spectra are maximally surface sensitive and intensities are dominated by scattering events occurring within the top $TiO_2$ layer, as portrayed in the inset to Fig. 5. Data were taken over a $200^o$ range which encompassed (100), (010) and ($\bar{1}00$) and were symmetry averaged. That is, intensities from $\phi = 0^o$ (100) to $\phi = 90^o$ (010) were averaged with those from $\phi = 90^o$ (010) to $\phi = 180^o$ ($\bar{1}00$) to remove spurious intensity variations arising from surface defects. Comparison reveals a strong resemblance between the F 1s and O 1s scans. This result is expected if F substitutes for O in the top layer, but is not expected if F is bound within a disordered surface phase, in which case the F 1s azimuthal scan would be featureless. F substitution for O is reasonable based on ionic radii for $F^-$ (1.33Å in VI-fold coordination environments) and $O^{2-}$ (1.40Å in VI-fold coordination environments) [20]. Based on F1s and O1s intensities averaged across the (100) → (010) quadrant at $\theta = 10^o$ to remove diffraction modulations, eqns. 1 and 2 evaluated at $\theta = 10^o$ yield a F concentration of 12% of the anions in the surface layer, in good agreement with value obtained from the full polar scan in the (100) plane (14%).

The effect of surface F on the electronic structure of STO is shown in Fig. 6, where we compare the VB spectra of STO(001) with and without F. The F 2p level falls at the bottom of the VB, as seen by the new feature at ~9.5 eV in the spectra for STO with F. Thus, contrary to simple intuition, $F_O$ is not a donor in STO despite the fact that F has one more electron than O. The highest occupied orbital (F 2p) falls at the bottom of the VB, rather than near the Fermi level. The fact that F is not a donor is also consistent with the fact that the carrier density, extracted from Hall effect measurements, is in agreement with the nominal Nb concentration. Inasmuch as the F concentration within the top unit cell is more than an order of magnitude higher than the Nb concentration, F-derived donor activity would result in a measurably higher carrier concentration.

Because F is not a donor in STO, we expect that doping the STO surface with F would not automatically pin the Fermi level or result in degenerate doping at the surface, and such is indeed the case. To show this result, we extract the VB maximum (VBM) relative to the Fermi level from Ti $2p_{3/2}$, Sr $3d_{5/2}$ and O 1s binding energies for Nb-doped STO(001) after different treatments. Here we used the energy differences between the CL peak (taken from fits to Voigt functions) and the VBM ($E_v$), in order to track $E_v$ following the different preparations. The general formula is $E_v = E_{CL} - (E_{CL} - E_v)_{ref}$ where the reference surface is flat-band STO(001) resulting from heating at 575°C in a oxygen plasma, and $(E_{CL} - E_v)_{ref}$ is 455.95(3) eV for Ti 2p$_{3/2}$, 130.54(3) eV for Sr 3d$_{5/2}$, and 527.10(3) eV for O 1s [21]. Based on the carrier concentrations measured by the Hall effect (mid $10^{19}$ cm$^{-3}$) [22], we estimate that in the bulk, the Fermi level is only a few hundredths of an eV below the conduction band minimum (CBM). Moreover, the STO bandgap is 3.2 eV [23]. Thus, a value of $E_v$ equal to ~3.15 eV corresponds to a flat-band condition at the surface. The results are shown in Table 1. The bands bend upward by several tenths of an eV after oxygen plasma cleaning at ambient temperature, and downward by a few tenths of an eV after vacuum annealing following oxygen plasma cleaning. The surface is in a flat-band state following annealing in the oxygen plasma, and then bend upward after a second HF etch, followed by oxygen plasma cleaning at ambient temperature. However, the F concentration remains unchanged within experimental error throughout these treatments. From these results, it is clear that factors other than F substitution for O control the surface state density and the associated band bending. These factors could include cation vacancy creation at levels below the detection limit of XPS, which may also be one means by which the surface electrically compensates F$^-$ substitution for O$^{2-}$.

Unless the F is chemically removed from the surface during heterojunction formation, it will remain at the interface and may effect the transport properties. The largest effect is likely to be a reduction in electron mobility associated with interface conductivity due to impurity scattering. Therefore, in order to obtain the best mobilities, it is of considerable interest to prepare STO(001) substrates in such a way that there is no F. We suggest that replacing the HF etch step with boiling in DI H$_2$O may be represent an improvement on the HF etch approach. Boiling for 30 minutes in DI H$_2$O rather than etching in HF reduces the F concentration by a factor of ~6, as discussed above. The surface morphology after the tube furnace anneal is quite flat as judged by AFM, and but does not exhibit the characteristic terrace-step structure that we and others observe on HF-etched STO. Rather, step bunching occurs. Moreover, this surface is $TiO_2$-termimated, as revealed by a polar scan of the Ti $2p_{3/2}$ and Sr $3p_{3/2}$ intensities in the (100) azimuth, seen in Fig. 7a. The Sr $3p_{3/2}$ and Ti $2p_{3/2}$ intensities exhibit diffraction modulation characteristic of the A and B sites in the perovskite lattice, respectively. Accordingly, the Ti $2p_{3/2}$ to Sr $3p_{3/2}$ peak area ratio at low take-off angles should be sensitive to the surface termination. Indeed, this ratio rises as $\theta$ drops, as qualitatively expected if the surface is $TiO_2$ terminated. Moreover, the monotonic behavior at low $\theta$ is reasonably well reproduced by a simple inelastic scattering simulation using equations similar to eqns. 1 & 2, but with parameters appropriate for the Ti $2p_{3/2}$ to Sr $3p_{3/2}$ core levels, and in which a $TiO_2$-terminated surface was modeled. This comparison is shown in Fig. 7b. Although these results are promising, it remains to be seen how the quality of epitaxial films grown on this surface compares to that for HF-etched substrates.

## 4. Conclusions

In summary, we show that F is unintentionally doped into O sites within the surface $TiO_2$ layer of STO(001) as a result of etching in HF. The doping level is of the order of 13 % of the anion sites of the top layer on average, and is not found in deeper layers. F doping does not result in either Fermi-level pinning or degenerate surface doping because the F 2p orbitals incorporate at the bottom of the valence band rather than at or near the Fermi level. The presence of F at the STO surface may have unintended consequences at buried interfaces involving STO. However, F contamination of the surface can be significantly reduced by replacing the HF etch with boiling in DI $H_2O$.

**Acknowledgements**

The authors thank Prof. Paul Lyman for a critical reading of the manuscript. This work was supported by the U.S. Department of Energy, Office of Basic Energy Sciences, Division of Materials Sciences and Engineering under Award number 10122, and was performed in the Environmental Molecular Sciences Laboratory, a national science user facility sponsored by the Department of Energy's Office of Biological and Environmental Research and located at Pacific Northwest National Laboratory.

# References

[1]  W. Meevasana, P. D. C. King, R. H. He, S. K. Mo, M. Hashimoto, A. Tamai, P. Songsiriritthigul, F. Baumberger, Z. X. Shen, Nat. Mat. 10 (2011) 114.

[2]  A. F. Santander-Syro, O. Copie, T. Kondo, F. Fortuna, S. Pailhes, R. Weht, X. G. Qiu, F. Bertran, A. Nicolaou, A. Taleb-Ibrahimi, P. Le Fevre, G. Herranz, M. Bibes, N. Reyren, Y. Apertet, P. Lecoeur, A. Barthelemy, M. J. Rozenberg, Nature 469 (2011) 189.

[3]  M. Kawasaki, K. Takahashi, T. Maeda, R. Tsuchiya, M. Shinohara, O. Ishiyama, T. Yonezawa, M. Yoshimoto, H. Koinuma, Science 266 (1994) 1540.

[4]  G. Koster, B. L. Kropman, G. Rijnders, D. H. A. Blank, H. Rogalla, Appl. Phys. Lett. 73 (1998) 2920.

[5]  J. Zhang, D. Doutt, T. Merz, J. Chakhalian, M. Kareev, J. Liu, L. J. Brillson, Appl. Phys. Lett. 94 (2009) 092904.

[6]  L. Qiao, T. C. Droubay, T. Varga, M. E. Bowden, V. Shutthanandan, Z. Zhu, T. C. Kaspar, S. A. Chambers, Phys. Rev. B 83 (2011) 085408.

[7]  J. J. Yeh and I. Lindau, At. Data Nucl. Data Tables 32 (1985) 1.

[8]  D. M. Tench and D. O. Raleigh, Natl. Bureau of Standards Special Publication 455 (1975) 229.

[9]  K. Szot, W. Speier, U. Breuer, R. Meyer, J. Szade, R. Waser, Surf. Sci. 460 (2000) 112.

[10] D. Kobayashi, H. Kumigashira, M. Oshima, T. Ohnishi, M. Lippmaa, K. Ono, M. Kawasaki, H. Koinuma, J. Appl. Phys. 96 (2004) 7183.

[11] R. P. Vasquez, Surf. Sci. Spectra 1 (1992) 24.

[12] C. S. Fadley, Prog. Surf. Sci 16 (1984) 275.

[13] S. A. Chambers and T. J. Irwin, Phys. Rev. B 38 (1988) 7858.

[14] S. A. Chambers and V. S. Sundaram, Appl. Phys. Lett. 57 (1990) 2342.

[15] S. A. Chambers and V. S. Sundaram, J. Vac. Sci. Technol. B 9 (1991) 2256.

[16] S. A. Chambers and T. T. Tran, Phys. Rev. B 47 (1993) 13023.

[17] T. T. Tran and S. A. Chambers, J. Vac. Sci. Technol. B 11 (1993) 1459.

[18] S. A. Chambers, Y. Gao, Y. J. Kim, M. A. Henderson, S. Thevuthasan, S. Wen, K. L. Merkle, Surf. Sci. 365 (1996) 625.

[19] S. A. Chambers, Advances in Physics 40 (1991) 357.

[20] R. D. Shannon, Acta Cryst. A 32 (1976) 751.

[21] S. A. Chambers, M. H. Engelhard, V. Shutthanandan, Z. Zhu, T. C. Droubay, L. Qiao, P. V. Sushko, T. Feng, H. D. Lee, T. Gustafsson, E. Garfunkel, A. B. Shah, J.-M. Zuo, Q. M. Ramasse, Surf. Sci. Rep. 65 (2010) 317.

[22] G. Y. Sun, C. Capan, S. A. Chambers, unpublished.

[23] K. van Bentham, C. Elsasser, R. H. French, J. Appl. Phys. 90 (2001) 6156.

Table 1 - Valence band maximum relative to the Fermi Level for HF-etched and tube-furnace annealed Nb:SrTiO$_3$(001) extracted from core-level binding energies

| surface treatment | $E_v$ from Ti 2p (eV) | $E_v$ from Sr 3d (eV) | $E_v$ from O 1s (eV) | average $E_v$ (eV) |
|---|---|---|---|---|
| O plasma clean – 2x10$^{-5}$ T, 30 min | 2.85(4) | 2.81(4) | 2.85(4) | 2.84(4) |
| Vac. anneal – 575°C 10 min. | 3.45(4) | 3.36(4) | 3.45(4) | 3.42(4) |
| O plasma clean while heating | 3.15(4) | 3.11(4) | 3.15(4) | 3.14(4) |
| 2$^{nd}$ HF etch – O plasma clean | 2.90(4) | 2.86(4) | 2.90(4) | 2.89(4) |

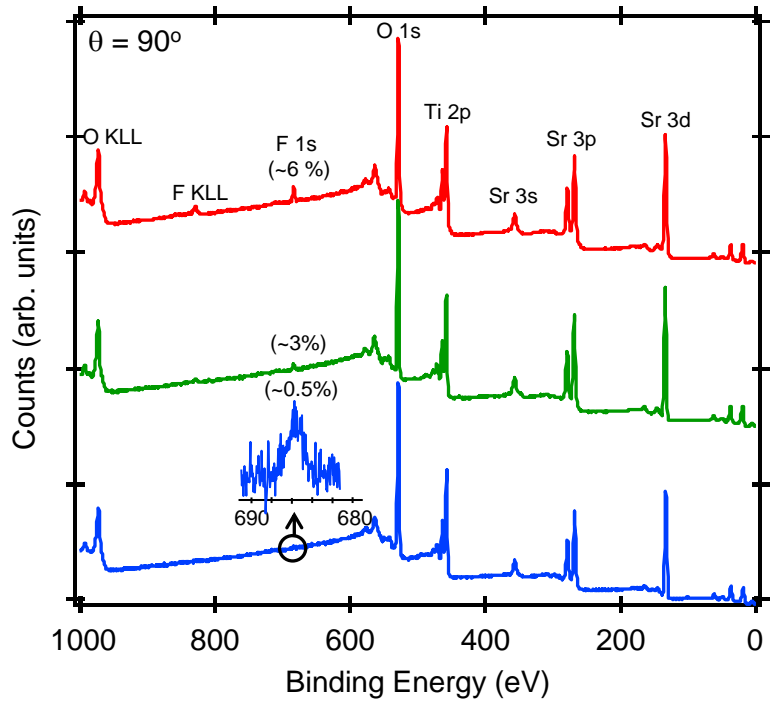

Fig. 1 – Normal emission (θ = 90°) XPS survey scans for three Nb-doped STO(001) surfaces that were etched in HF acid and tube furnace annealed in air. The top spectrum represents the highest F level we have observed (~6% of the anions within the probe depth, which is ~5 nm). The middle spectrum represents the average F level measured (~3%), and the sample in the bottom spectrum (~0.5%) was boiled in deionized water, rather than being etched in HF.

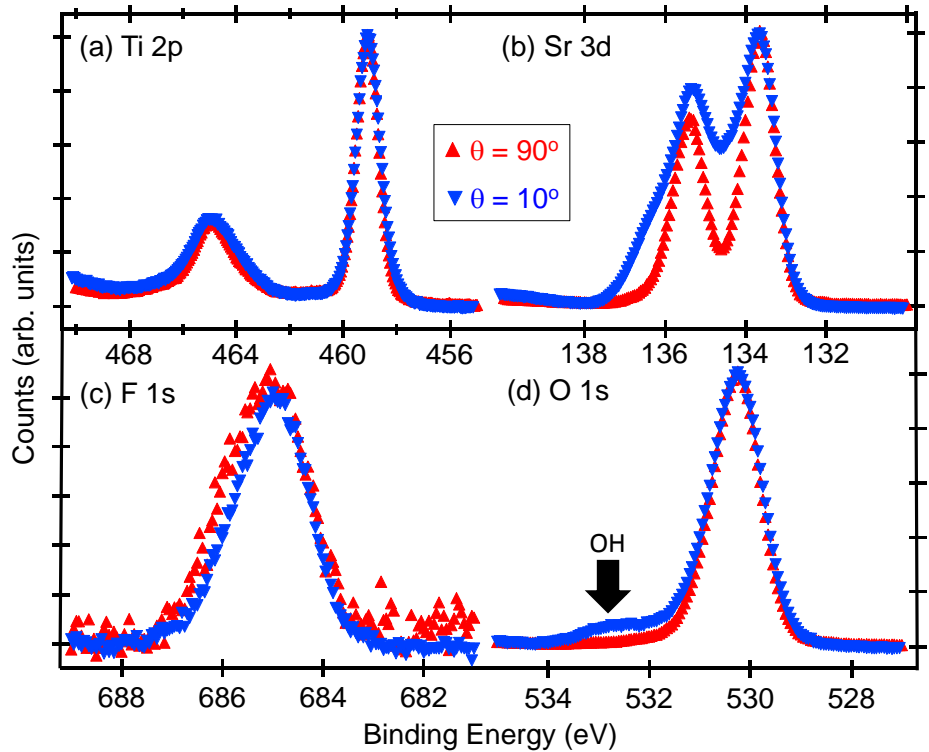

Fig. 2 – High-resolution core-level spectra measured at take-off angles (θ) of 90° and 10°, for which the approximate probe depths are 5 nm and 1 nm, respectively, for a Nb:STO(001) surface with ~3% F within the top ~5 nm.

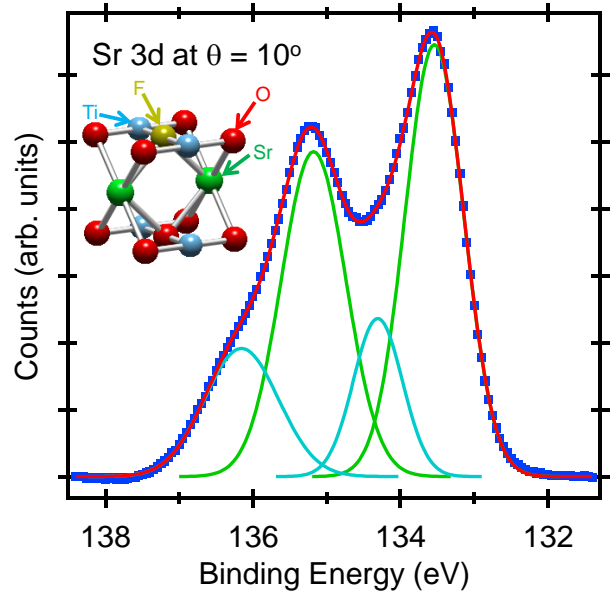

Fig. 3 – A fit of the θ = 90° Sr 3d spectrum from Fig. 2, showing the primary STO lattice peaks (more intense) and two "impurity" peaks, which we assign at least in part to second-layer Sr cations bound to F anion(s) in the surface layer, as depicted in the inset.

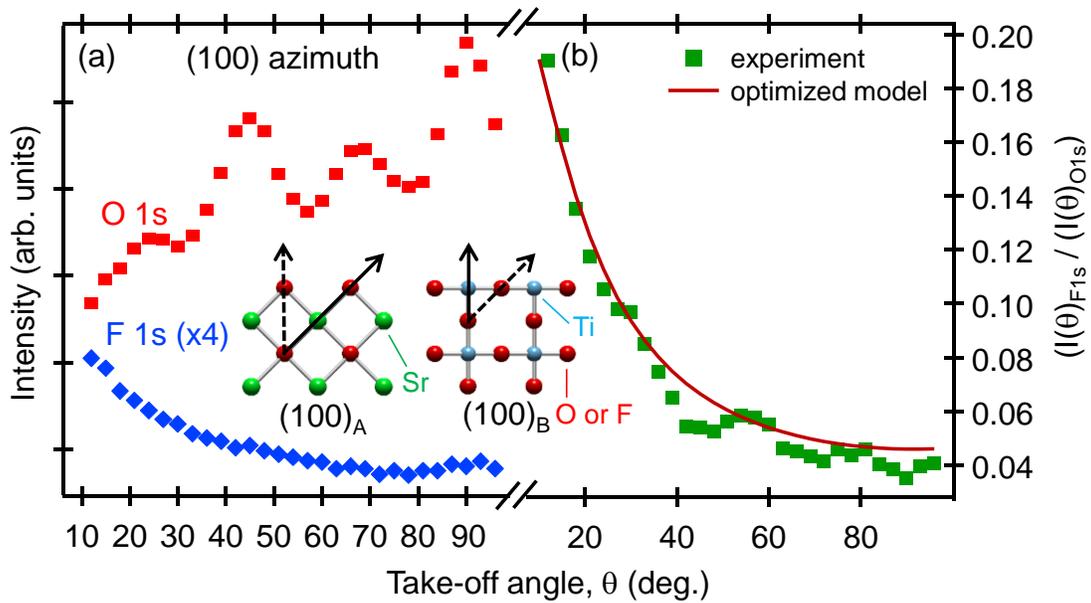

Fig. 4 – O 1s and F 1s polar scans in the (100) azimuth (a), along with the F 1s-to-O 1s peak area ratio, and a model calculation generated using eqns. 1 & 2, with $x_{F,1} = 0.14$ and $x_{F,2} = x_{F,3} = \ldots = 0$ (b).

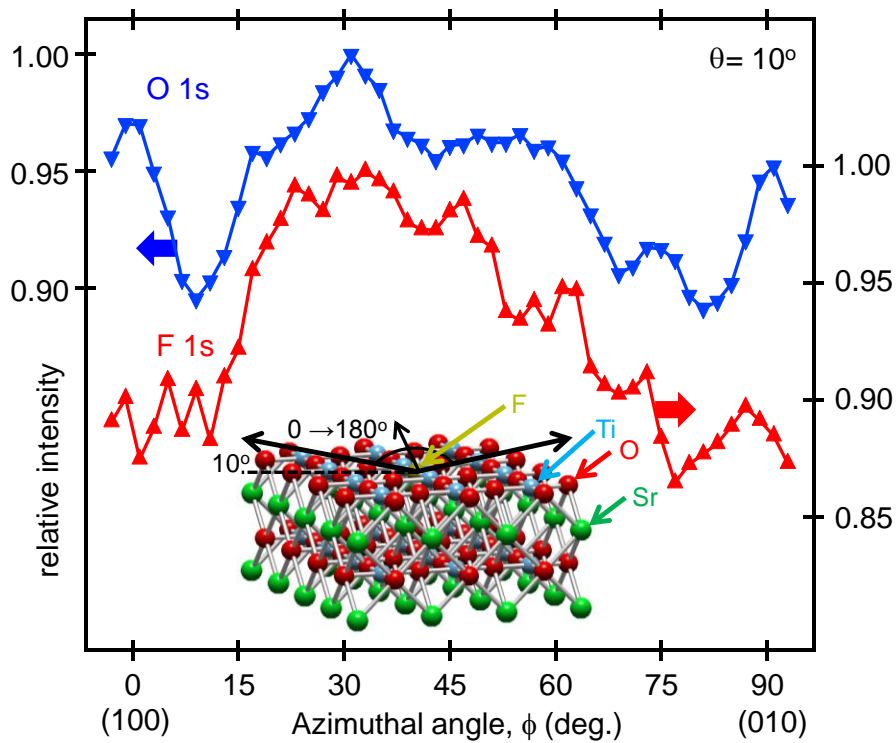

Fig. 5 – O 1s and F 1s azimuthal scans at $\theta =$ of $10°$ along with a structural diagram showing the full angular range over which the data were collected.

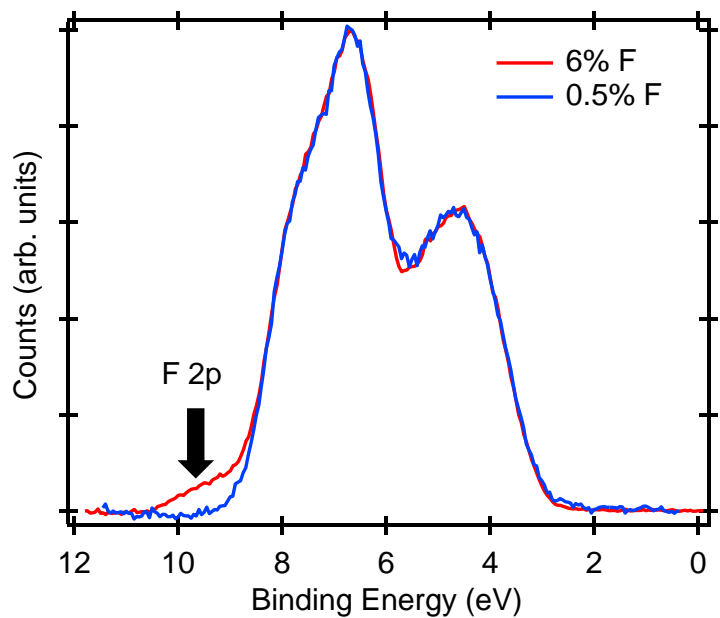

Fig. 6 – Valence band XPS for the STO(001) with and without F. The F 2p-derived density of states falls at the bottom of the valence band, as indicated by the arrow.

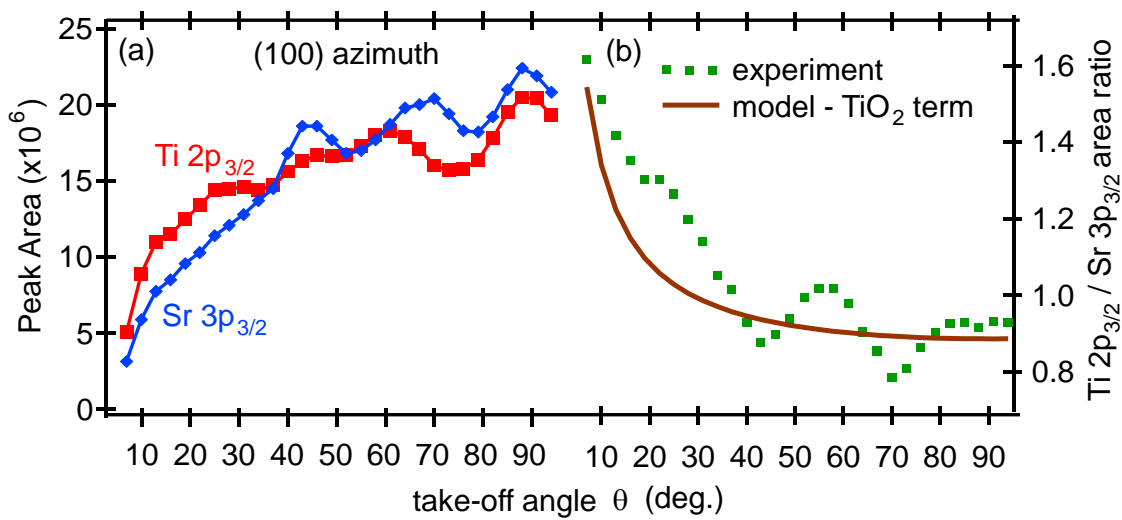

Fig. 7 – Ti $2p_{3/2}$ and Sr $3p_{3/2}$ polar scans in the (100) azimuth (a), along with the Ti $2p_{3/2}$ to Sr $3p_{3/2}$ peak area ratio, and a model simulation assuming a $TiO_2$-terminated surface (b).

18